\documentclass[draft,11pt]{article}
\usepackage{amssymb}
\newcommand{\nothing}[1]{}

\newcommand{\beq}[1]{\begin{equation}\label{#1}}
\newcommand{\eeq}{\end{equation}}

\newcommand{\sopr}[1]{\widetilde {\overline{#1}}}
\newcommand{\eq}{\triangleq}

\renewcommand{\emptyset}{\varnothing}

\newcommand{\varlimsup}{\mathop{\overline{\lim}}\limits}

\newcommand{\z}{{\textbf{\textit{z}}}}
\newcommand{\y}{{\textbf{\textit{y}}}}
\newcommand{\x}{{\textbf{\textit{x}}}}
\newcommand{\n}{{\textsf{\textit{n}}}}
\newcommand{\p}{{\textsf{\textit{p}}}}

\newcommand{\N}{{\cal N}}
\newcommand{\E}{{\cal E}}

\renewcommand{\O}{{\cal O}}
\newcommand{\X}{{\textbf{\textit{X}}}}
\newcommand{\A}{{\textbf{\textit{A}}}}
\newcommand{\B}{{\textbf{\textit{B}}}}

\renewcommand{\b}{{\textbf{\textit{b}}}}

\renewcommand{\u}{{\textbf{\textit{u}}}}
\renewcommand{\v}{{\textbf{\textit{v}}}}
\renewcommand{\S}{{\cal S}}
\renewcommand{\P}{{\cal P}}

\renewcommand{\l}{\ell}

\begin{document}

\begin{center}
{\Large\bf  Lectures on DNA Codes}~\footnote[0]{This article is a preprint
of the following paper (published in Russian along with its official version in English): 
{\em A.~Dyachkov, P.~Vilenkin, I.~Ismagilov, R.~Sarbayev, A.~Macula, D.~Torney,  P.~White}, 
"On dna codes"// Problems of Information Transmission, vol. 41, no. 4, pp. 349-367, 2005.}
\end{center}

 \begin{center}
   \textbf{ Arkadii G. D'yachkov}\\

\vspace{1cm}

Department of Probability Theory, Faculty of Mechanics and
Mathematics,\\
Moscow State University, Moscow 119992, Russia.\\
E-mail:\quad agd-msu@yandex.ru
\end{center}

\begin{abstract}

For  $q$-ary $n$-sequences, we develop the concept
of similarity  functions that can be used (for $q=4$) to model
a thermodynamic  similarity on  DNA
sequences. A similarity function is identified by
the length of
a longest common subsequence between two $q$-ary $n$-sequences.
Codes based on similarity functions are called
DNA  codes~\cite{d03,m04,d02}.
DNA codes are important components  in  biomolecular
computing~\cite{a94} and other biotechnical applications that
employ DNA hybridization assays.
The main aim of the given lecture notes -- to discuss
lower bounds on  the rate of optimal DNA codes for
a biologically motivated~\cite{m04} similarity function
called a block similarity  and for the conventional deletion
similarity function~\cite{lev65,lev74,lev01} used
in the theory of error-correcting codes.
We also present  constructions of suboptimal DNA codes based on
the parity-check code detecting one error in the Hamming
metric~\cite{ms77}.

\end{abstract}


\section{Introduction and Biological Motivation}
\qquad
Single strands of DNA are, abstractly,
$(A,C,G,T)$-quaternary sequences, with the four letters
denoting the respective nucleic acids: \textit{adenine} ($A$),
\textit{citosine} ($C$), \textit{guanine} ($G$),
and \textit{thymine}~($T$). Strands of DNA
sequence are oriented; for instance, $X=AACG$ is distinct
from $Y=GCAA$.  Furthermore, DNA is ordinarily double
stranded:  each sequence $X$, or strand, occurs with its
{\em reverse complement} $X'$, with reversal denoting that
the sequences of the two strands are oppositely oriented,
relative to one other, and with complementarity denoting
that the allowed pairings of letters, opposing one another
on the two strands, are $(A,T)$ or $(C,G)$---the canonical
Watson-Crick pairings. For instance, two sequences $X=AACG$
and $X'=CGTT$ are reverse complement of one another.
Obviously, for any strand $X$, we have~$\left(X'\right)'=X$.

Whenever two, not necessarily complementary, oppositely
directed DNA strands "mirror" one another, they are capable
of coalescing into a DNA duplex which is based on hydrogen
bonds between some pairs of nucleic acids.
Namely,  pair $(A,T)$ forms \textit{two} bonds,  pair
$(C,G)$ forms \textit{three} bonds, and any other pair
is called a \textit{mismatch} because it does not form any bond.
The process of forming DNA
duplexes from single strands is referred to as DNA
{\em hybridization}. The greatest energy of DNA
hybridization (the greatest stability of DNA duplex)
is obtained when the two sequences are
reverse complement of one another and the DNA duplex formed
is a Watson-Crick (WC) duplex. However, there are many
instances when the formation of non-WC duplexes are
energetically favorable.  The energy of DNA  hybridization
(the stability of DNA duplex)
$\E(X,Y)$  of two single DNA strands $X$ and $Y$ is, to a
first approximation, measured by the longest length of a
common subsequence (not necessary contiguous) of either
strand and the reverse complement of the other~\cite{d03}.
For two mutually reverse complementary strands $X$ and $X'$ of
length $n$, this measure plainly equals their length $n$,
i.e., the maximum number of Watson-Crick bonds
(complementary letter pairs) which may be formed between
two oppositely oriented strands:
$$
\E(X,X')=\max\limits_Y\,\E(X,Y')\,=\,
\max\limits_Y\,\E(Y',X)\,=\,\E(X',X)\,=\,n.
\eqno(1.1)
$$
For instance, if $X=AACG$ and $X'=CGTT$, then
$\E(X,X')\,=\,4$.

A DNA code $\X$ is a collection of $N$ single stranded DNA
sequences (codewords) of fixed length $n$ where each strand occurs with
its  reverse complement and no strand in the code equals
its  reverse complement~\cite{d03,d02}, i.e., if $X\in\X$,
then $X'\in\X$ and $X'\ne X$. In DNA hybridization assays,
the general rule is that formation of WC duplexes is good,
but the formation of non-WC duplexes is bad. A primary
goal of DNA code design is to be assured that a fixed
temperature can be found that is well above the melting
point of all non-WC duplexes and well below the melting
point of all WC duplexes that can form from strands in the
code. Thus the formation of any WC duplex must be
significantly more energetically favorable than all
possible non-WC duplexes.
 Note~\cite{d03} that for  biotechnical applications, the code length $n$,
$10\le n\le 40$, is experimentally accessible and that codes with
up to $N=10^9$ codewords could soon be called~for.
\bigskip

The following practical issue was an origin for the concept of DNA
code. Assume that we have $p$ types of some \textit{molecular
objects} and $p$~\textit{pools}. Each pool contains many identical
copies (clones) of the corresponding object. We need to perform an
experiment over all these pools. Since each experiment is expensive we
are interested in the \textit{junction} of these pools into one big
\textit{metapool} and performing only one experiment over this metapool.
Then we face a problem of singling out some copies of each object from
this mixture for analyzing experiment results.

For this purpose, there exists a method in which
codewords of a DNA code $\X$ of size $N$, where $N=2p$ is an even
number, are used as \textit{tags}. We fix  any $p$ codewords
$X(1),\ldots,X(p)$ of  $\X$ which are called \textit{capture tags} and the
corresponding reverse complementary codewords
$X'(1),\ldots,X'(p)$ called \textit{address tags}. Modern
technologies allow to generate many copies of each tag and mark each
molecular object by the corresponding tag. Then a metapool is created
and an experiment is performed. We assume that these processes do not
change capture tags.

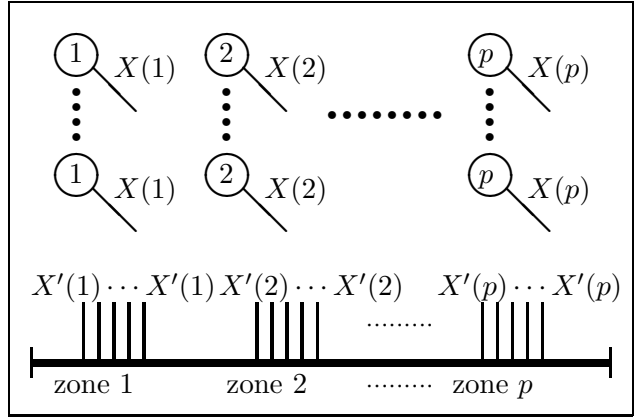
\begin{figure}
\begin{center}
\begin{picture}(87,57)
{\linethickness{1mm}
\put(5,7){\line(1,0){77}}
}
\put(5,5){\line(0,1){4}}
\put(82,5){\line(0,1){4}}
\put(2,0){\line(1,0){83}}
\put(2,0){\line(0,1){55}}
\put(85,55){\line(0,-1){55}}
\put(85,55){\line(-1,0){83}}
\thicklines
\put(2,0){
\multiput(10,7)(2,0){5}{\line(0,1){8}}
\put(3,16){$X'(1)\cdots X'(1)$}
\put(6,3){zone $1$}
}
\put(25,0){
\multiput(10,7)(2,0){5}{\line(0,1){8}}
\put(5,16){$X'(2)\cdots X'(2)$}
\put(6,3){zone $2$}
}
\put(55,0){
\multiput(10,7)(2,0){5}{\line(0,1){8}}
\put(4,16){$X'(p)\cdots X'(p)$}
\put(6,3){zone $p$}
}
\multiput(50,12)(1,0){9}{\circle*{0.5}}
\multiput(50,4)(1,0){9}{\circle*{0.5}}

\put(0,0){
\put(10,30){
\put(0,1){1}
\put(1,2){\circle{6}}
\put(3.24,0.24){\line(1,-1){5.66}}
\put(6,-1){$X(1)$}
}
\put(10,46){
\put(0,1){1}
\put(1,2){\circle{6}}
\put(3.24,0.24){\line(1,-1){5.66}}
\put(6,-1){$X(1)$}
}
\multiput(11,37)(0,2){4}{\circle*{1}}
}
\put(20,0){
\put(10,30){
\put(0,1){2}
\put(1,2){\circle{6}}
\put(3.24,0.24){\line(1,-1){5.66}}
\put(6,-1){$X(2)$}
}
\put(10,46){
\put(0,1){2}
\put(1,2){\circle{6}}
\put(3.24,0.24){\line(1,-1){5.66}}
\put(6,-1){$X(2)$}
}
\multiput(11,37)(0,2){4}{\circle*{1}}
}
\put(55,0){
\put(10,30){
\put(-0.5,1){$p$}
\put(1,2){\circle{6}}
\put(3.24,0.24){\line(1,-1){5.66}}
\put(6,-1){$X(p)$}
}
\put(10,46){
\put(-0.5,1){$p$}
\put(1,2){\circle{6}}
\put(3.24,0.24){\line(1,-1){5.66}}
\put(6,-1){$X(p)$}
}
\multiput(11,37)(0,2){4}{\circle*{1}}
}
\multiput(45,40)(2,0){8}{\circle*{1}}
\end{picture}
\caption{a metapool with capture tags $X(i)$ and address tags $X'(i)$}\label{pool}
\end{center}
\end{figure}

After this a solid support is taken. It is divided into $p$ separated
zones. Many copies of an address tag $X'(i)$ are immobilized
onto the corresponding $i$-th zone that physically segregates them.
Then the support is placed into the metapool.
This process is illustrated on Fig.~\ref{pool}.

Each pair of DNA sequences (codewords of DNA code $\X$)
in a pool may form a duplex except immobilized address tags.
In particular, any capture tag $X(i)$
may form a duplex with an address tag $X'(j)$. In this case,
the corresponding object of the $i$-th type finds itself settled on
the $j$-th zone of the support. Since there are many copies of each
object and many copies of each address tag, one can finally find
any type of object settled on $j$-th zone for any $j=1,\ldots,p$.

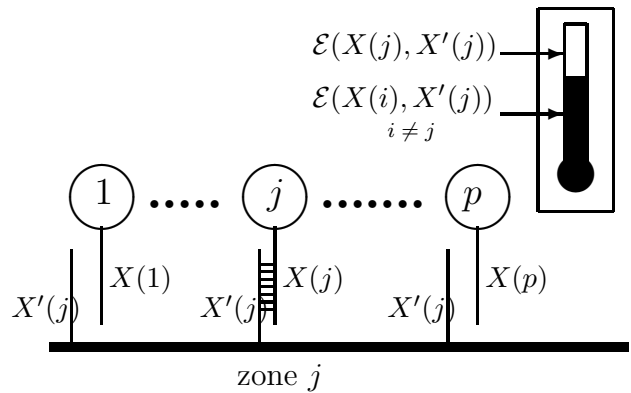
\begin{figure}
\begin{center}
\begin{picture}(87,52)
{\linethickness{1.2mm}
\put(5,7){\line(1,0){77}}
}
\put(30,2){\large zone $j$}
{\thicklines
\put(0,0){
\put(8,7){\line(0,1){13}}
\put(0,11){$X'(j)$}
\put(12,10){\line(0,1){13}}
\put(12,27){\circle{8}}
\put(11,26){\Large$1$}
\put(13,15){$X(1)$}
}
\multiput(19,26)(2,0){5}{\circle*{1}}
\put(25,0){
\put(8,7){\line(0,1){13}}
\put(0,11){$X'(j)$}
\put(10,10){\line(0,1){13}}
\put(10,27){\circle{8}}
\put(9,26){\Large$j$}
\put(11,15){$X(j)$}
}
\multiput(42,26)(2,0){7}{\circle*{1}}
\put(50,0){
\put(8,7){\line(0,1){13}}
\put(0,11){$X'(j)$}
\put(12,10){\line(0,1){13}}
\put(12,27){\circle{8}}
\put(10,26){\Large$p$}
\put(13,15){$X(p)$}
}
\multiput(33,12)(0,1){7}{\line(1,0){2}}
\put(75,30){\circle*{5}}
{\linethickness{3mm}
\put(75,30){\line(0,1){13}}
}
\put(70,25){\line(1,0){10}}
\put(70,25){\line(0,1){27}}
\put(80,52){\line(-1,0){10}}
\put(80,52){\line(0,-1){27}}
\put(73.5,30){\line(0,1){20}}
\put(76.5,30){\line(0,1){20}}
\put(73.5,50){\line(1,0){3}}
\put(65,38){\vector(1,0){8.5}}
\put(65,46){\vector(1,0){8.5}}
\put(40,46){$\E(X(j),X'(j))$}
\put(40,39){$\E(X(i),X'(j))$}
\put(50,35){\scriptsize$i\ne j$}
}
\end{picture}
\caption{a separation of the $j$-th objects}\label{separation}
\end{center}
\end{figure}

Let a stability function $\E$ expresses the melting temperature
of a duplex. Assume that for an index $j\in\{1,2,\ldots,p\}$
a certain temperature range separates  large value $\E(X(j),X'(j)$
from  small values $\E(X(i),X'(j)$ for $i\ne j$
and small values $\E(X(i),X(j))=\E\left(X(i),(X'(j))'\right)$ for any $i$ and~$j$.
This means that there exists a temperature range at which all duplexes
on the $j$-th zone melt except those which are formed by $X(j)$ and $X'(j)$.
Finally, only the objects of the $j$-th type will be settled on the
corresponding zone and that separates them from the other types,
see Fig.~\ref{separation}. Whenever this condition holds for all values $j$,
we are able to separate all types of objects.
\bigskip

\bigskip

The mathematical analysis of  DNA hybridization is based
on the concept of similarity functions that can  be used to
model a thermodynamic similarity  on single stranded  DNA
sequences. For two quaternary $n$-sequences $X$ and $Y$,
the longest length of a sequence occurring  as a (not
necessary contiguous) subsequence of both is called a
deletion similarity $S^{\lambda}(X,Y)$ between $X$ and~$Y$.
We supposed~\cite{d03,d02} that the deletion similarity
$S^{\lambda}(X,Y)$ identifies the number of base pair
bonds in a hybridization assay between $X$ and the reverse
complement of~$Y$, i.e., the energy of DNA hybridization
$\E(X,Y')$ satisfying $(1.1)$ is defined as follows
$$
\E(X,Y')\,=\,\E(X',Y)
\,=\,S^{\lambda}(X,Y)\,=\,S^{\lambda}(Y,X).
\eqno(1.2)
$$

Let $D$, $1\le D\le n-1$, be a fixed integer.
A DNA code $\X$ is called a  DNA code of distance $D$
based on deletion similarity or, briefly, an
$(n,D)$-code~\cite{d03,d02} if the deletion similarity
$$
S^{\lambda}(X,Y)\le n-D-1\qquad
\mbox{for any}\quad
X,\,Y\in\X,\quad Y\ne X.
\eqno(1.3)
$$
Definition $(1.2)$ and condition $(1.3)$ mean that
the energy of DNA hybridization
$$
\E(X,Y')\le n-D-1 \qquad
\mbox{for any}\quad
X,\,Y\in\X,\quad Y\ne X,
$$
i.e., any strand $X\in\X$ and the reverse
complement of the other strand $Y\in\X$ can never form $\ge n-D$
base pair bonds in a hybridization assay.
In the theory of deletion - correcting codes, condition $(1.3)$,
by itself, specifies  codes capable to correct any
combination of $D$ deletions~\cite{lev65}.

\textbf{Example 1.1}\quad DNA code $\X=\{X,X',Y,Y'\}$, where
$$
X=ACAT,\quad X'=ATGT,
\qquad
Y=ATAC,\quad Y'=GTAT,
\eqno(1.4)
$$
is a $(n,D)$-code of length $n=4$ and distance $D=1$
because $n-D-1=2$ and sequence $Z=AT$ of length~$2$ is the
longest common subsequence between any pair of strands in
DNA code~$\X$. Hence,
$$
\E(X,X)=\E(X',X')=S^{\lambda}(X,X')=2,
\qquad
\E(Y,Y)=\E(Y',Y')=S^{\lambda}(Y,Y')=2,
$$
$$
\E(X,Y)=\E(X',Y')=S^{\lambda}(X,Y')=2,
\qquad
\E(X,Y')=\E(X',Y)=S^{\lambda}(X,Y)=2.
$$

In paper~\cite{m04}, we introduced the concept of common
block subsequence, namely: a common subsequence $Z$ of
sequences $X$ and $Y$ is called a common block subsequence
if any two consecutive elements of $Z$ which are
consecutive in $X$ are also consecutive in $Y$ and vice
versa.  For two quaternary  $n$-sequences $X$ and $Y$, the
longest length of a sequence occurring  as a common block
subsequence of both is called a block similarity between
$X$ and~$Y$. For example, sequence $Z=AT$ of length~$2$ is
the longest common block subsequence between any pair of
strands in DNA code~$(1.4)$. Thus, DNA code~$(1.4)$ can be
considered as DNA $(4,1)$-code based on block similarity.

The first conventional issue of coding theory~\cite{ms77}
for DNA codes -- to get a lower random coding bound on the
rate of DNA codes and, hence, to identify values of the
distance fraction $D/n$  for which DNA code size grows
exponentially when $n$ increases. The given problem is more
difficult than the corresponding problem for
deletion - correcting codes.  For instance, we cannot apply
the best known random coding bounds~\cite{dan94} on the
rate of deletion-correcting codes because these bounds were
proved for codes which are not invariant under the reverse
complement transformation.
The second  conventional issue of coding theory for DNA
codes -- to present constructions of DNA codes.
The aim of our lecture notes is to discuss  bounds and constructions
for  DNA codes based on the deletion and block similarities which have
a good biological motivation to model a thermodynamic
similarity on DNA sequences~\cite{m04}.
We will study $q$-ary DNA codes which are
generalizations of quaternary
DNA codes.


\section{Notations, Definitions and Examples}
\qquad The symbol $\eq$ denotes definitional equalities and the
symbol $[n]\eq\{1,2,\dots,n\}$ denotes the set of integers from~1
to~$n$. Let  $q=2,4,\dots$ be a fixed even integer,
$\A\eq\{0,1,\dots,q-1\}$ be the standard  alphabet of size
$|\A|=q$ and $\lfloor u\rfloor$ $\,(\lceil u\rceil)$ denote the
largest (smallest) integer~$\le u\,$~$(\ge u)$.
Introduce the  binary entropy function
$$
h_q(u)\eq-u\log_qu-(1-u)\log_q(1-u),\quad 0<u<1.
$$

Consider two
arbitrary $q$-ary $n$-sequences
$$
\x=(x_1,x_2,\dots,x_n)\in \A^n,\quad
\y=(y_1,y_2,\dots,y_n)\in \A^n.
$$
In what follows, we will denote by symbol $S=S(\x,\y)$ an
arbitrary symmetric function satisfying conditions
$$
0\le S(\x,\y)=S(\y,\x)\le S(\x,\x)\,=\, n,\,\,\,
\x\in \A^n,\,\,\, \y\in \A^n,
\eqno(2.1)
$$
and called~\cite{d03} a {\em similarity} function.
For instance, an {\em additive} similarity function
$$
S^{\alpha}(\x,\y)\eq\sum\limits_{i=1}^n\,S^{\alpha}(x_i,y_i),\quad
\mbox{where}\quad
S^{\alpha}(x,y)\eq\cases{1 & if $x=y$,\cr
                      0  & if $x\ne y$,\quad $x,y\in \A$,  \cr}
$$
is the number of positions in which $\x$ and $\y$ coincide.
Function $S^{\alpha}(\x,\y)$ can be called the Hamming similarity because
$
n\,-\,S^{\alpha}(\x,\y)
$
is the well-known Hamming distance function (metric) applied
in the theory of error-correcting codes~\cite{ms77}.

Let $\l\in[n]$ and  $m=1,2,\dots,\l$.
By symbol
$$
\z=(z_1,z_2,\dots,z_{\l})\in \A^{\l},
\quad\mbox{where}\quad
z_m=x_{i_m}=y_{j_m},
$$
$$
1\le i_1<i_2<\cdots<i_{\l}\le n,\quad
1\le j_1<j_2<\cdots<j_{\l}\le n,
$$
we will denote a {\em common subsequence} of length
$|\z|\eq \l$ between $\x$ and~$\y$. By definition, the {\em
empty} subsequence of length $|\z|\eq 0$ is a common
subsequence between any sequences $\x$ and~$\y$.

\textbf{Definition 1.}~\cite{lev65}.
Let $S^{\lambda}(\x,\y)$, $0\le S^{\lambda}(\x,\y)\le n$, denote the length
$|\z|$ of {\em longest} common subsequence~$\z$ between sequences $\x$ and~$\y$.
The number $S^{\lambda}(\x,\y)$  is called
a {\em deletion similarity} between $\x$ and~$\y$.
Evidently, the function $S^{\lambda}=S^{\lambda}(\x,\y)$
satisfies~(2.1).

\textbf{Definition 2.}~\cite{m04}.
A common subsequence
$
\z=(z_1,z_2,\dots,z_{\l}),\; 2\le\l\le n,
$
is called a {\em common block subsequence} of
length $|\z|\eq \l$ between $\x$ and $\y$ if any two
consecutive elements $z_m,z_{m+1}$, $m=1,2,\dots,\l-1$,
which are  consecutive (separated) in $\x$ are also
consecutive (separated)  in
$\y$ and vice versa, i.e,
$$
\left(z_m=x_{i_m},z_{m+1}=x_{i_m+1}\right)
\leftrightarrow
\left(z_m=y_{j_m},z_{m+1}=y_{j_m+1}\right).
$$
By definition, any common subsequence $\z$ of length
$|\z|=0$ or $|\z|=1$ is a common block subsequence.
Let $S^{\beta}(\x,\y)$,  $0\le S^{\beta}(\x,\y)\le n$,
denote the length $|\z|$ of {\em longest} sequence occurring
as a common block subsequence $\z$ between sequences $\x$ and~$\y$.
The number $S^{\beta}(\x,\y)$ is called a
{\em block similarity} between  $\x$ and~$\y$.
Obviously, $S^{\beta}=S^{\beta}(\x,\y)$
satisfies~(2.1) and
$$
S^{\beta}(\x,\y)\,\le\, S^{\lambda}(\x,\y),\,\,\,
\x\in \A^n,\,\,\, \y\in \A^n.
$$

\textbf{Definition 3.}~\cite{d03,d02}.
If $q=2,4,\dots$, then
$$
\bar{x}\eq\,(q-1)-x,\quad x\in \A=\{0,1,\dots,q-1\},
$$
is called a {\em complement} of a letter $x$.
For sequence
$\x=(x_1,x_2,\dots,x_{n-1},x_n)\in \A^n$,
we define its {\em reverse  complement}
$\sopr{\x}\eq
(\bar{x}_n,\bar{x}_{n-1},\dots,\bar{x}_{2},\bar{x}_{1})\in \A^n$.
Obviously, if $\y\eq\sopr{\x}$, then $\x=\sopr{\y}$ for any $\x\in\A^n$.
If $\x=\sopr{\x}$, then $\x$ is called a {\em self reverse
complementary} sequence. If $\x\ne\sopr{\x}$, then a pair $(\x\,,\,\sopr{\x})$
is called a {\em pair of mutually reverse complementary}
sequences.
\medskip

\textbf{Example 2.1.}\quad Let $q=2$, $n=8$ and
$$
\x=(0,1,0,1,1,0,1,1),\quad
\y=(0,0,1,0,0,1,1,0).
$$
Obviously, $S^{\alpha}(\x,\y)=2$.
The deletion similarity $S^{\lambda}(\x,\y)=6$
because $6$-sequence
$$
\z\eq(0,1,0,1,1,0)=(x_1,x_2,x_3,x_4,x_5,x_6)=(y_2,y_3,y_4,y_6,y_7,y_8)
$$
is the longest sequence occurring as a common subsequence between $\x$
and~$\y$. The block similarity  $S^{\beta}(\x,\y)=5$
because sequence
$$
\z\eq(0,1,0,1,0)=(x_1,x_2,x_3,x_5,x_6)=(y_2,y_3,y_4,y_7,y_8)
$$
is the longest sequence occurring as a common block subsequence between $\x$
and~$\y$.

\textbf{Example 2.2.} Let $q=2$, $n=10$ and
$$
\x=(0,1,1,0,0,0,1,1,1,1),\quad
\y\eq\tilde{\bar{\x}}=(0,0,0,0,1,1,1,0,0,1)
$$
be a pair of mutually reverse complementary sequences.
We have $S^{\alpha}(\x,\y)=4$.
The deletion similarity $S^{\lambda}(\x,\y)=S^{\lambda}(\x,\tilde{\bar{\x}})=8$
because the self reverse complementary sequence
$$
\z\eq(0,0,0,0,1,1,1,1)=\tilde{\bar{\z}}=
(x_1,x_4,x_5,x_6,x_7,x_8,x_9,x_{10})=
(y_1,y_2,y_3,y_4,y_5,y_6,y_7,y_{10})
$$
is the longest sequence occurring as a common subsequence between $\x$
and~$\y\eq\tilde{\bar{\x}}$. The  block similarity
$S^{\beta}(\x,\y)=S^{\beta}(\x,\tilde{\bar{\x}})=6$
because the following self reverse complementary sequence
$$
\z\eq(0,0,0,1,1,1)=\tilde{\bar{\z}}=(x_4,x_5,x_6,x_7,x_8,x_9)=
(y_2,y_3,y_4,y_5,y_6,y_7)
$$
is a longest sequence occurring as a common block subsequence between $\x$
and~$\y=\tilde{\bar{\x}}$.
\medskip

Let $\x(1),\x(2),\dots,\x(N)$, where
$
\x(k)\eq(x_1(k),x_2(k),\dots,x_n(k)),\;
x_i(k)\in \A,\; k\in[N],
$
be {\em codewords} of a
$q$-ary {\em code}~$\X=\{\x(1),\x(2),\dots,\x(N)\}$
of {\em length} $n$ and  {\em size} $N$, where $N=2,4,\dots$
be an {\em even} number.
Let $D$, $1\le D\le n-1$, be an arbitrary integer.
\medskip

\textbf{ Definition 4.}~\cite{d03,m04,d02}.
A code $\X$ is called a DNA
{\em $(n,D)$-code based on similarity function}
$S=S(\x,\y)$ (briefly, $(n,D)$-code) if the following two conditions are fulfilled.
$(i)$~For any number $k\in[N]$ there exists
$k'\in[N]$, $k'\ne k$, such that $\x(k')=\sopr{\x(k)}\ne\x(k)$.
In other words, $\X$ is a collection of $N/2$ pairs of mutually
reverse complementary sequences.
$(ii)$~For any $k,k'\in[N]$, where $k\ne k'$, the similarity
$S(\x(k),\x(k'))\le n-D-1$. We will also say that code $\X$ is a
 DNA code of  length $n$, {\em distance}~$D$
and {\em similarity}~$n-D-1$.
\medskip

For $q=4$,  a biological motivation of  $(n,D)$-codes
 based on deletion similarity $S^{\lambda}=S^{\lambda}(\x,\y)$
 was suggested in~\cite{d03}. If only condition $(ii)$ is
 retained, then an $(n,D)$-code based on deletion similarity is
 a  code of length $n$ capable to correct any combination
 of $\le D$ deletions~\cite{lev65}. A biological motivation of
quaternary  DNA codes based on block similarity
$S^{\beta}=S^{\beta}(\x,\y)$ was suggested in~\cite{m04}.
\medskip

For given  $n$ and $D$, we denote by $N_q(n,D)$  the
{\em maximal  size} of  $(n,D)$-codes.
If $d$, $0<d<1$, is a fixed number, then
$$
R_q(d)\eq\varlimsup_{n\to\infty}\frac{\log_qN_q(n,\lfloor dn\rfloor)}{n}
\eqno(2.2)
$$
is called a {\em rate} of~$(n,\lfloor dn\rfloor)$-codes.
\medskip

We will use notations with upper indices
$N^{\lambda}_q(n,D)\,,\,R^{\lambda}_q(d)$ and
$N^{\beta}_q(n,D)\,,\,R^{\beta}_q(d)$
for the corresponding parameters of DNA codes based on
similarity functions $S^{\lambda}$ and~$S^{\beta}$.
From inequalities between considered similarity functions
it follows that
$N^{\lambda}_q(n,D)\le N^{\beta}_q(n,D)$ and
$R^{\lambda}_q(d)\le R^{\beta}_q(d)$.
\medskip

\textbf{Remark 2.1.}
If $D=1,2\dots$ is fixed and $n\to\infty$, then
$$
N^{\lambda}_q(n,D)\le\;\frac{D!}{(q-1)^D}\cdot
\frac{q^n}{n^{D}}\cdot(1+o(1)).
\eqno(2.3)
$$
This upper bound follows from the corresponding
results~\cite{lev65,ten84,lev02} (see, also~\cite{lev74}, p.~272)
obtained for codes capable to
correct any combinations of $\le D$ deletions.
\medskip

 \textbf{Remark 2.2.} One can easily understand that the conventional
Hamming bound on the size of block codes with distance $D+1$
is a trivial upper bound on  $N^{\beta}_q(n,D)$, i.e.,
$$
N^{\beta}_q(n,1)\le q^{n-1},\qquad
N^{\beta}_q(n,D)\le q^n\left/\sum_{i=0}^{\lfloor
D/2\rfloor}\,{n\choose i}\cdot(q-1)^i\right.,\quad D\ge2.
$$

For $D=1$  an
improvement of this trivial bound is given by
\medskip

\textbf{Theorem 2.1.}\quad {\em The maximal size}
$N^{\beta}_q(n,1)\le \left(q^{n-1}+q\right)/2$.
\medskip

\textbf{Proof of Theorem 2.1.} Consider an arbitrary
$q$-ary DNA code $\X=\{\x(k), k\in[N]\}$
of length $n$, distance~$D=1$ and block similarity~$n-2$.
For each codeword $\x(k)$, there exists one or two
 tail subsequences of length $n-1$ obtained by  deletions of the first or
the last element of~$\x(k)$. Let  $\X$ contain $N_1$ $(N_2=N-N_1)$  codewords
which yield  one (two) tail subsequences of length~$n-1$.
Obviously, $N_1\le q$. From item $(ii)$ of  Definition~4, it follows
that there are $N_1+2N_2$ distinct tail subsequences of length~$n-1$. Thus
one can write $\;N_1 + 2N_2 \leq q^{n-1},\; N_1\leq q.\;$
These two inequalities lead to
$
N =N_1 + N_2 \leq \frac{q^{n-1} + q}{2}.
$

Theorem 2.1 is proved.
\medskip

\textbf{Example 2.3.} If $q=2$ and $n=4$, then a DNA code of
length $n=4$,  size $N=4$,  distance $D=1$
 and block (deletion) similarity
$n-D-1=2$ contains 2 pairs of mutually reverse complementary
codewords:
$\,{\bf 0000 \; 1111}$ and $\,{\bf 0110 \; 1001}$.
Obviously, from Theorem~2.1 it follows that the given code has the maximal
size and $N^{\beta}_2(4,1)=N^{\lambda}_2(4,1)=4$.

\section{Suboptimal  DNA Codes for Distance $D=1$}
\qquad
In this section, we  assume that $n$ is a number divisible
by~$q$, where $q=2,4,\dots$ is an {\em even} number. Hence, $n$ is an
{\em even} number as well. We also remind that the {\em
complement}
of a letter $a\in\A\eq\{0,1,\dots,q-1\}$ is defined as~$\bar
a\eq(q-1)-a\,\in\A$. Therefore, $\bar a\ne a$ for any~$a\in\A$.
We say that a codeword
$\x\in\A^n$ satisfies the parity-check condition if the
arithmetic sum of its elements is a number divisible
by~$q$. Let $M_q(n)$ denote the set of all these codewords:
$$
M_q(n)\,\eq\,\{\x=(x_1,x_2,\dots,x_n)\in\A^n\,:\,x_1+\cdots+x_n\equiv 0
\,(\mathop{\mbox{mod}} q)\},\quad
|M_q(n)|=q^{n-1}.
\eqno(3.1)
$$
Any subset $T\subseteq M_n(q)$ is called a \textit{parity-check
code}. The set $M_n(q)$ is the optimal code of size $q^{n-1}$ detecting one error
in the Hamming metric~\cite{ms77}. It is called the
\textit{maximal parity-check code}.
We will construct suboptimal DNA codes for distance $D=1$ which are
{\em subcodes} of~$M_q(n)$.
Obviously, for each codeword $\x\in M_q(n)$, its reverse complement
$\sopr{\x}\in M_q(n)$.

\subsection{Formulations of Results}
\quad
In Sect.~3.2, we prove

\textbf{Theorem 3.1.}
 {\em There exists a $q$-ary DNA code of length $n$,
distance  $D=1$,  block similarity $n-D-1=n-2$ and
size}
\begin{itemize}
\item
$N=\frac12\,\left(q^{n-1}+q\right)$\quad {\em if}\quad $\;n=qk,\;$ $k=1,3,5,\dots$;
\item
$N=\frac12\,\; q^{n-1}$ \quad {\em if}\quad  $q = 2^m$, $n=2^{m+k}$,  $\;k\ge1$;
\item
$N\ge\frac12\,\left(q^{n-1}-\frac{q^{n/2+1}-1}{q-1}\right)$
\quad {\em if}\quad  $\;n=qk,\;$ $k=2,4,6,\dots$.
\end{itemize}
\medskip

\textbf{Remark 3.1.} If $n=qk$, where $k=1,3,5,\dots$ is an arbitrary odd number,
then Theorem~2.1 means that the construction of Theorem~3.1 is optimal.
If $q$ is fixed and $n\to\infty$, then Theorem~2.1 means
that the construction of Theorem~3.1 is asymptotically optimal.
\medskip

\textbf{Example 3.1.} For $n=q=4$, the construction of
optimal  DNA code  from Theorem~3.1 is illustrated by the following table
which contains $4^{3}=64$ codewords satisfying
the parity-check condition, namely: for each codeword, the
sum of its elements is a number divisible by~$4$.
$$
\begin{array}{llll}
\par\vspace*{0.2cm}\par
\underline{{\bf 0000,3333}}  &   &  & \\
  {\bf 0013,0233} & {\bf3001,2330} &
  \underline{{\bf1300,3302}}  & {\bf0130,3023}\\
{\bf 0031,2033}  &  {\bf1003,0332} &
{\bf3100,3320}  &  {\bf0310,3203}\\
\underline{{\bf 0103,0323}} &  {\bf3010,3230} &
\par\vspace*{0.2cm}\par
\underline{{\bf0301,2303}} &  {\bf1030,3032}\\
{\bf 0112,1223} & {\bf2011,2231} &
 \underline{{\bf1201,2312}} & {\bf1120,3122}\\
{\bf0121,2123}  & {\bf1012,1232} &
{\bf2101,2321}  & {\bf1210,3212} \\
\par\vspace*{0.2cm}\par
\underline{{\bf0211,2213}}  & {\bf1021,2132} &
\underline{{\bf1102,1322}}  & {\bf2110,3221} \\
\underline{{\bf 0022,1133}} &  {\bf2002,1331} &
  \underline{{\bf2200,3311}} &  {\bf0220,3113}\\
\par\vspace*{0.2cm}\par
{\bf0202,1313}  & {\bf2020,3131} & & \\
\underline{{\bf1111,2222}} & & &
\end{array}
$$
These
codewords are written as $\frac12\cdot4^{3}=32$ pairs of
mutually reverse complementary codewords. Any row of the table consists
of 1, 2, or 4 pairs. In any row, the first (second)
codewords are obtained as consecutive left (right) cyclic
shifts of the first (second) codeword of any fixed pair of
the row.  If we eliminate from the table all 15
pairs from the second and
fourth  columns of the table, then one can easily check that
the rest 17 mutually reverse complementary pairs will constitute
a quaternary DNA code
$\X$ of length $n=4$, size $N=2\cdot17=34$, block
distance~$D=1$ and  block similarity~$n-D-1=2$.
 We mark by the symbol
"underline"  pairs of codewords (there are 10 such pairs)
from code $X$ which have pairwise  deletion
similarities~$\le2$.  They constitute a quaternary DNA
code of length~$n=4$, size~$N=2\cdot10=20$, deletion
distance~$D=1$ and deletion similarity~$n-D-1=2$.  This
means that the maximal size~$N_4^{\lambda}(4,1)\ge20$.
\medskip

In Sect.~3.3, we prove

\textbf{Theorem 3.2.}
{\em Let $n=qk$, where $q=2,4,\dots$ is an even number
and $k=1,3,\dots$ is an odd number.
Let there exists a parity-check code $T$,
correcting single deletions, i.e., $T\subset M_n(q)$ and
the deletion similarity
$\S^\lambda(\x,\y)\le n-2$ for any $\x,\y\in T$,~$\x\ne\y$.
Then there exists a DNA $(n,1)$-code $T'\subset M_n(q)$ of
size $|T'|\ge|T|$.}
\medskip

We will use the following  construction~\cite{ten84}
of a a parity-check code $T$ correcting single deletions.
{\bf a)} Consider a partition of the set
$\A^n$ into $q$ subsets $\,M^1(\beta)$, $\;\beta=0,1,\ldots,q-1$, where
$$
M^1(\beta)\,\eq\,\left\{\x=(x_1,x_2,\dots,x_n)\in\A^n\,:\,
x_1+\cdots+x_n\equiv\beta\,(\mathop{\mbox{mod}}q)\right\}
$$
In particular, the maximal parity-check code~$M_n(q)=M^1(0)$.
{\bf b)} For each $\x\in\A^n$, we introduce a binary sequence
$(\alpha_2,\ldots,\alpha_n)$, where
$$
\alpha_i\eq\cases{
1 & if  $\;x_i\ge x_{i-1}$,\cr
0, & if $\;x_i < x_{i-1},\quad i=2,3\dots,n$.\cr}
$$
Consider a partition of the set
$\A^n$ into $n$ subsets $\,M^2(\gamma)$, $\;\gamma=0,1,\ldots,n$, where
$$
M^2(\gamma)\,\eq\,\left\{\x=(x_1,x_2,\dots,x_n)\in\A^n\,:\,
\sum\limits_{i=2}^n\,(i-1)\,\alpha_i\,\equiv\,\gamma
\;(\mathop{\mbox{mod}}n)\right\}.
$$
{\bf c)}
The intersection of  two partitions defined in items~{\bf a)}
and~{\bf b)}  yields a partition of the
set $\A^n$ into  $nq$ subsets having the form
$T(\beta,\gamma)\eq M^1(\beta)\cap M^2(\gamma)$.
One can prove~\cite{ten84} that every  subset of this partition
is a code correcting single deletions. Hence,
the size of a maximal code correcting single deletions
exceeds~$q^n/(nq)=q^{n-1}/n$.

If we fix $\beta=0$, then we obtain a partition of the set
$M_n(q)$ into $n$ subsets of the form $T(0,\gamma)$, $0\le\gamma\le n-1$.
Each of these subsets can be applied  as  a parity-check code $T$ for
Theorem~3.2.
If we choose a  code having the maximal size
$$
|T|\,=\,\max\limits_{0\le\gamma\le n-1}\; |T(0,\gamma)|\,\ge\,
\frac{|M_n(q)|}{n}\,=\,\frac{q^{n-1}}{n},
$$
then we obtain the following lower bound on
the maximal size of DNA~$(n,1)$-code.
\medskip

\textbf{Corollary.}
{\em  If $n=qk$, where $k=1,3,\dots$ is an odd number, then}
$$
N_q^{\lambda}(n,1)\ge\frac{q^{n-1}}{n}.
$$
\medskip

\textbf{Example 3.2.} One can easily check that the following
collection containing $11$ pairs of mutually reverse complementary codewords:
$$
\begin{array}{llll}
\;{\bf 0000 \quad  3333} \quad & \;{\bf 1111 \quad 2222} \quad &
\;{\bf 0022 \quad  1133} \quad & \;{\bf 2200 \quad 3311} \\
\;{\bf 0330 \quad  3003} \quad & \;{\bf 1221 \quad 2112} \quad &
\;{\bf 0011 \quad  2233} \quad & \;{\bf 1100 \quad 3322} \\
\;{\bf 0120 \quad  3123} \quad & \;{\bf 1301 \quad 2302} \quad &
\;{\bf 0231 \quad  2013} \quad & \;
\end{array}
$$
is a quaternary DNA code of length~$n=4$,
size~$N=22$, deletion distance~$D=1$ and
deletion similarity~$n-D-1=2$.
Note that only the first 4 pairs
satisfy the parity check condition~(3.1).

\textbf{Example 3.3.} One can also easily check that the collection
of 24 codewords:
$$
\begin{array}{llll}
\;{\bf 0000 \quad  3333} \quad & \;{\bf 1111 \quad 2222} \quad &
\;{\bf 0022 \quad  1133} \quad & \;{\bf 2200 \quad 3311} \\
\;{\bf 0321 \quad  2103} \quad & \;{\bf 2012 \quad 1231} \quad &
\;{\bf 0011 \quad  2233} \quad & \;{\bf 1100 \quad 3322} \\
\;{\bf 3013 \quad  0230} \quad & \;{\bf 0033 \quad 3300} \quad &
\;{\bf 1122 \quad  2211} \quad & \;{\bf 1302 \quad 3120}
\end{array}
$$
is a quaternary  code of length~$n=4$,
size~$N=24$, deletion distance~$D=1$ and
deletion similarity~$n-D-1=2$. This code is a code capable to
correct single deletions. The given code is not a DNA code because
the last~6 codewords of this code are self reverse
complementary sequences.

\textbf{Remark 3.2.}
One can prove that codes from Examples~3.2 and~3.3 are optimal
codes, i.e., their sizes~$N=22$ and~$N=24$ are maximal possible
for the corresponding codes of length~$n=q=4$.
Proofs of these statements are omitted here because they
are awkward and we do not know any generalizations for codes of
length~$n>4$.

\subsection{Proof of Theorem 3.1}
\quad
The following important property of the maximal
parity-check code $M_q(n)$ takes place.

\textbf{Lemma 3.1.}\quad
{\em If $n=qk$, where $k=1,3,\dots$,
then code $M_q(n)$ does not contain self reverse complementary
codewords}.

\textbf{Proof of Lemma 3.1.} By contradiction. Let there exist a
codeword
$$
\x=(x_1,x_2,\dots,x_n)=\sopr{\x}\in M_q(n).
$$
Then
$\;x_{n-i+1}=q-1-x_i,\;$  $\,i=1,2,\dots,n/2,\;$
and the sum
$$
\sum\limits_{i=1}^n\,x_i\,=\,
\sum\limits_{i=1}^{n/2}\,[x_i+(q-1-x_i)]\,=
\,\frac{n}{2}(q-1)=\frac{q k}{2}(q-1)
$$
is a number divisible by $q$.
This contradicts to the condition~$k=1,3,\dots$.

Lemma 3.1 is proved.

For any sequence $\x\in\A^n$, we define its first {\em left cyclic shift} $\;T_1,\;$
i.e.,
$$
T_1(\x)\,\eq\,(x_2,x_3,\dots,x_n,x_1)\,\in\A^n \quad\mbox{if}\quad
\x=(x_1,x_2,\dots,x_n)\in\A^n.
$$
Introduce the $(k+1)$-th left cyclic shift $T_{k+1}$, $k=1,2,\dots$, i.e.,
$T_{k+1}(\x)\eq T_1(T_k(\x))$. By the similar way we define the $k$-th
{\em right} cyclic shift $T_k$, where~$k<0$.
Let symbol $T_0$  be the identity operator. For indices
$i,k\in[n]$, we define index $i+k\in[n]$ as the corresponding sum by
modulo~$n$. Obviously, the $i$-th element of $T_k(\x)$ has the
form~$T_k(\x)_i=x_{i+k}$.

The set $\O(\x)\eq\{T_k(\x)\,:\,k=0,1,\dots,n-1\}$
containing all cyclic shifts of $\x\in\A^n$
is called an {\em orbit generated by}~$\x$. Let $\l(\x)\eq|\O(\x)|$ denote
the orbit size. Note that $n$ is a number divisible
by~$\l\eq\l(\x)$. For any $\y\in\O(\x)$,
the orbit $\O(\y)=\O(\x)$, the size $\l(\y)=\l(\x)=\l$, the
$\l$-th shift~$T_{\l}(\y)=\y$ and~$\O(\x)=\{T_k(\x)\,:\,k=0,1,\dots,\l-1\}$.

In addition, it is easy to see that
$$
\sopr{T_k(\x)}=T_{-k}(\sopr{\x}), \quad \x\in\A^n, \quad k=1,2,\dots.
\eqno(3.2)
$$
It means that the set $\{\sopr{\y}\,:\,\y\in\O(\x)\}$ is an orbit
generated by $\sopr{\x}$. Thus, we obtain a {\em reverse complement
operator} for orbits. If an orbit
$\O(\x)$ does not contain self reverse complementary  sequences, then
$\O(\x)\cap\O(\sopr{\x})=\emptyset$ and for any $\y\in\O(\x)$,
its reverse complement~$\sopr{\y}\in\O(\sopr{\x})$.
The given orbits $\O(\x)$ and $O(\sopr{\x})$ are
called  {\em mutually  reverse complementary orbits}.

If an orbit $\O(\x)$  contains a self reverse complementary sequence,
then $\O(\x)=\O(\sopr{\x})$ and $\O(\x)$ is
called a {\em self reverse complementary orbit}.
The following statement gives the structure of all  self reverse
complementary orbits.
\medskip

\textbf{Lemma 3.2.}
{\em If an orbit $\O(\x)=\O(\sopr{\x})$, then the orbit size $\l=\l(\x)$
is an even number and $\O(\x)$ contains exactly two
self reverse complementary sequences which are the $\,\l/2$-th cyclic
shifts of each other. In addition, if these two self reverse
complementary sequences $($without loss of generality$)$
are $\x$ and $T_{\l/2}(\x)$, then  the rest $\l-2$ sequences
from orbit $\O(\x)$ can be divided into $(\l-2)/2$ pairs of
mutually reverse complementary sequences of the form}
$$
\left(T_{\l/2-i}(\x)\,,\,T_{\l/2+i}(\x)\right),\quad
\mbox{{\em where}}\quad
T_{\l/2+i}(\x)=\sopr{T_{\l/2-i}(\x)},\;
i=1,2,\dots,(\l-2)/2.
\eqno(3.3)
$$

\textbf{Proof of Lemma 3.2.} As far as $\sopr{\x}\in\O(\x)$ then
there exists an integer $k$, $k=0,1,\dots,n-1$, for which the $k$-th cyclic
shift~$T_k(\x)=\sopr{\x}$. Hence, for any $i=1,2,\dots,n$, the
$i$-th coordinate of $\,T_k(\x)=\sopr{\x}\,$ is~$\,x_{i+k}=\overline{x_{n+1-i}}\,$.

Let $k$ be an odd number. Since $n$ is an even number we
put the integer~$i\eq\frac{n+1-k}{2}$. This  leads to equality
$x_{(n+1+k)/2}=\overline{x_{(n+1+k)/2}}$
which contradicts to the condition $\bar{a}\ne a$,~$a\in\A$.
Therefore, $k$ is an even number, i.e.,~$k=2t$.

Consider sequence $\y\eq T_t(\x)\in\O(\x)$.
Taking into account the above properties of $\x$, one can easily check that
the  $i$-th coordinate of $\y$ is
$$
(\y)_i\eq y_i=x_{i+t}=x_{(i-t)+2t}=\overline{x_{n+1-(i-t)}}=
\overline{x_{n+1-i+t}}=\overline{y_{n+1-i}}=(\sopr{\y})_i,
\quad i=1,2,\dots,n.
$$
We have $\y=\sopr{\y}$ and the $\l$-th shift $T_{\l}(\y)=\y=\sopr{\y}$
because $\y\in\O(\x)$. This means that the orbit size
$\l=\l(\x)=\l(\y)$ is an even number, i.e.,~$\l=2m$.

Let $\z$ be an arbitrary self reverse complementary sequence and
$\z=\sopr{\z}\in\O(\x)$. From (3.2) it follows
$$
\sopr{T_m(\z)}=T_{-m}(\sopr{\z})=T_{-m}(\z)=T_{m-\l}(\z)=T_m(T_{-\l}(\z))=
T_m(\z),\quad m=\l/2.
$$
On the other hand, let $s$ be an arbitrary integer such that
$T_s(\z)$ be a self reverse complementary sequence.
For any~$i\in[n]$, we obtain
$$
z_{i+2s}=z_{i+s+s}=(T_s(\z))_{i+s}=(\sopr{T_s(\z)})_{i+s}=
(T_{-s}(\sopr{\z}))_{i+s}=(T_{-s}(\z))_{i+s}=z_i,
$$
i.e., $T_{2s}(\z)=\z$. It follows that $2s$ is a number divisible by~$\l=2m$
and $s$ is a number divisible by~$m=\l/2$. Therefore, the orbit
$\O(\x)$ contains exactly two self reverse complementary sequences
$\y\eq T_t(\x)$ and~$T_{\l/2}(\y)$. The form (3.3) for
mutually reverse complementary sequences follows from~(3.2).

Lemma 3.2 is proved.
\medskip

\textbf{Lemma 3.3.} {\em For any codewords $\x,\y\in M_q(n)$, $\x\ne\y$,
the block similarity $\S^{\beta}(\x,\y)=n-1$ if and only if
either  $T_1(\x)=\y$ or~$T_{-1}(\x)=\y$.}

\textbf{Proof of Lemma 3.3}.
Let $\S^{\beta}(\x, \y)=n-1$.
Then $\x$ and $\y$ have a common block of length~$n-1$.
Each of these codewords has an extra symbol which is
either the first or the last symbol of the corresponding codeword.
From the parity-check condition it follows that this extra symbol
is the same in $\x$ and~$\y$ and, hence, the given symbol is the
first (last) symbol in $\x$ ($\y$) or vise versa. In other words,
$T_1(\x)=\y$ or~$T_{-1}(\x)=\y$. The converse statement is
evident.

Lemma 3.3 is proved.
\medskip

\textbf{Lemma 3.4.} {\em Let $\O(\x)=\O(\sopr{\x})\in M_q(n)$
and $\l=\l(\x)=4k$. Then there exists a subset $\X\subset\O(\x)$
of size $|\X|=2k$ which is a DNA code of block similarity~$\,n-2$.}

\textbf{Proof of Lemma 3.4.}
From Lemma~3.2 it follows that without loss of generality, we can assume  that
$\x=\sopr{\x}$. Define code
$$
\X\eq\{T_m(\x)\,:\, m=1,3,\dots\,\l-1\}.
$$
Obviously, the size  $|\X|=\frac12\cdot|\O(\x)|=\l/2=2k$ because
for any $\y\in\O(\x)$, the $s$-th shift $T_s(\y)=\y$ if and only
if $s$ is a number divisible by~$\l=4k$.
In virtue of Lemma~3.2 and equality $\l/2=2k$,
the set $\X$ does not contain self reverse complementary codewords.
From (3.3) it follows that for codeword
$\y=T_{\l/2-i}(\x)\in\X$,
codeword~$\sopr{\y}=T_{\l/2+i}(\x)\in\X$,
$i=1,3,\dots,(\l-2)/2$. Finally, Lemma~3.3 shows that the
block similarity of code~$\X$ does not exceed~$n-2$.

Lemma 3.4 is proved.
\medskip

We divide the set $M_q(n)$, $n=qk$, into four nonintersecting subsets
$G_i$, $i=1,2,3 ,4$.
Subset $G_1$ contains all orbits of size~$\l=1$.
Subset $G_2$ contains all self reverse complementary orbits of
size~$\l=2$.
Subset $G_3$ contains all self reverse complementary orbits of
size~$\l=4k$, $k=1,2\dots$.
Subset $G_4$ contains all other orbits. In virtue of Lemma~3.1,
$G_4$ consists of all pairs of mutually reverse complementary
orbits.
For some values $n=qk$, subset $G_2$ and (or) subset $G_3$
are empty.

Obviously, $G_1=\{\x=(a,a,\ldots,a),a\in\A\}$ and the size $|G_1|=q$.
The set $G_1$ is invariant under the reverse complement transformation
and does not contain self reverse complementary  codewords.
The block similarity between any two codeword from $G_1$ is equal
to zero. Therefore, $G_1$ satisfies DNA code definition.

{\sf1)} Let $n=qk$, $k=1,3,5,\dots$. In virtue of Lemma~3.1, the set
$M_q(n)$ does not contain self reverse complementary
codewords~$\x=\sopr{\x}$.
Hence,  $G_4$ contains $q^{n-1}-q$ codewords and
$G_4$ consists of mutually reverse complementary
orbits  $\O(\x)$ and $O(\sopr{\x})$.

We construct a required code $\X$ in the following way.
{\sf1a)} The set $G_1$ is included in~$\X$.
{\sf1b)} For each pair of mutually reverse complementary
orbits  $\O(\x)$ and $O(\sopr{\x})$, code $\X$ contains
one-half of their codewords having the following form:
$$
\left(T_k(\x)\,,\,T_{-k}(\sopr{\x}\right)\,:\,k=0,2,4,\dots,\l-1,
\quad \l=\l(\x)=\l(\sopr{\x}).
$$
Taking into account (3.2) and Lemma~3.3, it is easy to see that
the code $\X$ is a DNA code of block similarity~$n-2$.
The size of $\X$ has the form
$$
|\X|=q+\frac{q^{n-1}-q}{2}=\frac{q^{n-1}+q}{2}.
$$
\medskip

{\sf2)} Let $q=2^m$ and $n=2^{m+m'}$, $m'\ge1$. In this case,  $G_2$
contains self reverse complementary orbits of size~$\l=2$ and
codewords $\x\in G_2$ have the form
$$
G_2=\{\x\,:\, \x=(a,\bar a,a,\bar
a,\ldots,a,\bar a),\; a\in\A\}, \qquad |G_2|=q.
$$
Set $G_4$ consists of mutually reverse complementary
orbits  $\O(\x)$ and $O(\sopr{\x})$.

We construct a required code $\X$ in the following way.
{\sf2a)}~The set $G_1$ is included in~$\X$. {\sf2b)}~Elements of $G_2$
are not included in~$\X$. {\sf2c)}~Code $\X$ contains one-half of codewords
from the set~$G_3$ according to Lemma~3.3.
{\sf2d)}~Code $\X$ contains one-half of codewords
from the set~$G_4$ having the form
described in item~{\sf1b)}. Obviously, $\X$ is a DNA code of block similarity~$n-2$.
The size of $\X$ has the form
$$
|\X|=|G_1|+\frac{|G_3|+|G_4|}{2}=|G_1|+\frac{|M_q(n)|-|G_1|-|G_2|}{2}=
q+\frac{q^{n-1}-2q}{2}=q^{n-1}/2.
$$
\medskip

{\sf3)} Let $n=qm$, where $m,q=2,4,\dots$ be an arbitrary even numbers.
In this case, $n$ is a number divisible by $4$, i.e.~$n=4k$.
 Let
$M^1_q(n)\subset M_q(n)$ be subcode of
code $M_q(n)$, where $M^1_q(n)$ contains all orbits
$\O(\x)\subset M_q(n)$ of size $\l(\x)=n$.
For any  $\O(\x)\in M_q(n)\backslash M^1_q(n)$,
the size $\l(\x)\le n/2=2k$. Obviously, the total size of all orbits
$\O(\x)$ for which $\l(\x)=d$ does not exceed~$q^d$. This leads to
the inequality
$$
|M_q(n)\backslash M^1_q(n)| \leq \sum_{d=1}^{2k}q^d =
\frac{q^{2k+1} - 1}{q - 1}\quad\mbox{or}\quad
|M^1_q(n)|\ge q^{n-1}-\frac{q^{n/2+1}-1}{q-1}.
$$
For any  $\O(\x)\in M^1_q(n)$,
the size $\l(\x)=n=4k$. Therefore, according to the
construction described in item~{\sf1b)} and  Lemma~3.4, we obtain
a DNA code $\X$ of block similarity~$n-2$ and size
$$
|\X|\,\ge\,\frac12\,\left(q^{n-1}-\frac{q^{n/2+1}-1}{q-1}\right).
$$

Theorem 3.1 is proved.


\subsection{Proof of Theorem 3.2}
\quad
Let a sequence $\x\in\A^n$. We will say that an integer-valued vector
$$
\n(\x)\,=\,\n=\,(n_0,n_1,\ldots,n_{q-1}),\quad 0\le n_x\le n,\quad
x\in\A=\{0,1,2,\dots q-1\},
$$
is  a {\em composition} of  $\x$ if
$n_x$ is equal to the number of entries of the symbol $\,x\in\A\,$ in~$\x$.
The  reverse complement transformation of a sequence $\x$ leads to the
reverse transformation of its composition:
$\n(\sopr{\x})=\bar\n\eq(n_{q-1},\ldots,n_1,n_0)$.
In what follows, we will consider codewords $\x\in M_q(n)$ having
compositions $\n$ for which
$$
\sum_{x=0}^{q-1}n_x=n,\qquad \sum_{x=0}^{q-1}\,x\cdot n_x\equiv
0\;(\mathop{\mbox{mod}} q).\eqno (3.4)
$$

\textbf{Lemma 3.5.}\quad {\em If $\,\x,\y\in M_q(n)\,$ and
$\,\n(\x)\ne\n(\y)$, then} $\S^{\lambda}(\x,\y)\le n-2$.

\textbf{Proof of Lemma 3.5.} By contradiction.
Consider two arbitrary  codewords
 $\x,\y\in M_q(n)$ with deletion similarity
$S^{\lambda}(\x,\y)=n-1$. Obviously, these codewords can be
obtained by two distinct insertions of the same symbol into
their common subsequence of length~$n-1$. Therefore, $\x$ and $\y$
should have the same composition that contradicts to the condition
of Lemma~3.5.

Lemma~3.5 is proved.
\medskip

\textbf{Lemma 3.6.}\quad
{\em Let $n=qk$, where $k=1,3,\dots$ be an arbitrary odd number.
If  composition  $\n$ satisfies  $(3.4)$, then $\n\ne\bar\n$.
In particular, code $M_q(n)$ does not contain self reverse complementary
codewords}.

\textbf{Proof of Lemma 3.6.} By contradiction. Let there exist a
composition $\n$ for which~$\n=\bar\n$. It means that
$n_x=n_{q-1-x}$, $\; x\in\A\,$, and the sum
$
\sum\limits_{x=0}^{q-1}\,n_x\,=\,2\cdot\sum\limits_{x=0}^{q/2}\,n_x\,=\,n.
$
Hence,
$$
\sum\limits_{x=0}^{q-1}\,x\cdot n_x\,=\,
\sum\limits_{x=0}^{q/2}\,[x+(q-1-x)]\cdot n_x\,=\,
(q-1)\cdot\sum\limits_{x=0}^{q/2}\, n_x\,=\,\frac{(q-1)n}{2}\,
=\,\frac{(q-1)qk}{2}.
$$
In virtue of $(3.4)$, the right-hand side
is a number divisible by~$q$. This contradicts to~$k=1,3,\dots$.

Lemma~3.6 is proved.
\medskip

Let a subset $T\subset M_q(n)$ be a code correcting single
deletions, i.e., for any
codewords $\x,\y\in T$, $\x\ne\y$,
the deletion similarity~$\S^{\lambda}(\x,\y)\le n-2$.
We will prove that there exists a DNA code
 $T'\subset M_q(n)$, $|T'|\ge |T|$, having the same property.

Let $T$ be a fixed code correcting single
deletions. We choose a set of compositions
$\N$ satisfying (3.4) in the following way.
Consider all composition pairs $(\n,\bar\n)$
satisfying~(3.4).
In virtue of Lemma~3.6,
$\n\ne\bar\n$ and the set
$
M(\n)\eq\{\x\in M_q(n)\,:\,\n(\x)=\n\}
$
does not contain self reverse complementary
codewords.
For any pair $(\n,\bar\n)$ the set
 $\N$ contains exactly one element of the pair, namely: if
$|T\cap M(\n)|\ge|T\cap M(\bar\n)|$, then $\N$ contains $\n$,
and $\N$ contains $\bar\n$, otherwise.
Introduce the set
$$
M(\n,T)\eq T\cap M(\n)\subset M_q(n),\qquad
\sopr{M(\n,T)}\eq\{\sopr{\x}\,:\,\x\in M(\n,T)\}\subset M_q(n).
$$
Then the set
$$
T'\eq\bigcup_{\n\in\N}M(\n,T)\cup\sopr{M(\n,T)}
$$
is a DNA code of size~$|T'|\ge |T|$. From Lemma~3.5 it follows
that for any codewords  $\x,\y\in T'$ having distinct
compositions, the deletion similarity~$S^{\lambda}(\x,\y)\le n-2$.
From construction of  $T'$ it follows
that for any codewords  $\x,\y\in T'$ having the same composition,
we have $\x,\y\in T$ or~$\sopr{\x},\sopr{\y}\in T$.
And, therefore, in this case the deletion similarity is
$S^{\lambda}(\sopr{\x},\sopr{\y})=S^{\lambda}(\x,\y)\le n-2$.

Theorem~3.2 is proved.

\section{Bounds for DNA Codes}

\subsection{Formulations of results}
\qquad
Theorem~4.1 presents lower  bounds
on the size $N^{\lambda}_q(n,D)$ and rate $R^{\lambda}_q(d)$
of DNA codes based on deletion similarity.
Let $d=d^{\lambda}_q$, $\,0<d^{\lambda}_q<(q-1)/q$, be the unique
root of equation
$$
\frac{1+d}{2}=d\log_q(q-1)+h_q(d).
$$

\textbf{Theorem 4.1.}\quad
$(i)$. {\em If $D=1,2\dots$ is fixed and $n\to\infty$, then}
$$
N^{\lambda}_q(n,D)\ge\;\frac14\cdot D!^2\cdot\left(\frac{q}{(q-1)^{2}}\right)^D\cdot
\frac{q^n}{n^{2D}}\,\cdot\,(1+o(1)).
\eqno(4.1)
$$
$(ii)$. {\em If $0<d< d^{\lambda}_q$, then the rate $R^{\lambda}_q(d)>0$
and  the lower bound
$$
R^{\lambda}_q(d)\,\ge\,\underline{R}^{\lambda}_q(d)\eq
1+d-2[d\log_q(q-1)+h_q(d)], \qquad 0<d<d^{\lambda}_q,
\eqno(4.2)
$$
holds.}
\medskip

\textbf{Example 4.1.}  For the binary case,
$d^{\lambda}_2=0.13340$ and for the most important
quaternary case, $d^{\lambda}_4=0.27029$.
In addition, $d^{\lambda}_6=0.34902$ and
$d^{\lambda}_8=0.40324$.
\medskip

Theorem~4.2 gives lower  bounds on
the size $N^{\beta}_q(n,D)$  and rate $R^{\beta}_q(d)$
of DNA codes based on the similarity of blocks.
Let $v=v(d)$, $\;0<v(d)<d$,
be the unique root of equation
$$
\left(\frac{1-d}{v}-1\right)\left(\frac{d}{v}-1\right)^2=1,
\qquad 0<v<d<\frac12.
\eqno(4.3)
$$
One can easily understand that
 $v(d)$ is calculated using the following
recurrent method: $w_1\eq2$,
$$
w_{m+1}\,=\,1\,+\,\frac{1}{\sqrt{\frac{1-d}{d}\,w_m-1}},\quad
m=1,2,\dots,\quad
v(d)=\frac{d}{\lim\limits_{m\to\infty}w_m}.
$$
Define the function
$$
E_q(d)\,\eq\,(1-d)\,h_q\left(\frac{v(d)}{1-d}\right)\,+
\,2d\,h_q\left(\frac{v(d)}{d}\right),\quad 0<d<\frac12.
\eqno(4.4)
$$
Let $d_q^{\beta}$, $0<d_q^{\beta}\le1/2$, be the unique root of equation
$1-d=E_q(d)$.

\textbf{Theorem 4.2.}
\quad
$(i)$. \quad {\em If $D=1,2\dots$ is fixed and $n\to\infty$, then}
$$
N^{\beta}_q(n,D)\ge\;\frac14\cdot\frac{D!}{q^D}\cdot
\frac{q^n}{n^{D}}\,\cdot\,(1+o(1)).
\eqno(4.5)
$$
$(ii)$.\quad {\em If
$0<d<d_q^{\beta}$, then the rate $R_q^{\beta}(d)>0$ and
the following lower bound
$$
R_q^{\beta}(d)\,\ge\,\underline{R}_q^{\beta}(d)\,\eq\, (1-d)-E_q(d),
\quad 0<d<d_q^{\beta}.
\eqno(4.6)
$$
holds}.

Theorem~4.2 will be  proved in Sect.~4.3 with the help of
a random coding method described in Sect.~4.2.
We briefly present
the similar proof of Theorem~4.1 in Sect.~4.4.
\medskip

\textbf{Example 4.2.} \quad We calculated
$d_2^{\beta}=0.17888$, $d_4^{\beta}=0.35752$,
$d_6^{\beta}=0.44523$ and $d_8^{\beta}=1/2$.
It means that the critical points
for block similarity  exceed the corresponding
critical points (see, Example~4.1) for deletion similarity.
\medskip

\subsection{Random Coding Method for DNA Codes}
\quad
In this section, we develop a general random coding method for
DNA codes. Let $\S=\S(\x,\y)$ be an arbitrary similarity function~(2.1).
For integers $0\le s\le n$, we define two sets
$$
\P(n,s)\eq\{(\x,\y)\in\A^n\times\A^n\,:\,S(\x,\y)=s\},
\quad
\bar{\P}(n,s)\eq\{\x\in\A^n\,:\,S(\x,\tilde{\bar{\x}})=s\}.
\eqno(4.7)
$$
Consider two random sequences
$$
\u=(u_1,u_2,\dots,u_n),\qquad
\v=(v_1,v_2,\dots,v_n),
$$
with independent
identically distributed components  having the uniform
distribution on~$\A$. Obviously, the corresponding
probability distributions  of random variables $\S(\u,\v)$
and $\S(\u,\tilde{\bar{\u}})$ have the form:
$$
\Pr\{\S(\u,\v)=s\}=\frac{|\P(n,s)|}{q^{2n}},\quad
\Pr\{\S(\u,\tilde{\bar{\u}})=s\}=
\frac{|\bar{\P}(n,s)|}{q^{n}},\quad 0\le s\le n.
\eqno(4.8)
$$

A lower bound on $N_q(n,D)$
called a {\em random coding bound} is formulated as
\medskip

\textbf{Lemma 4.1.} {\em For any $D$, $1\le D\le n-1$, the number
$$
N_q\left(n,D\right)\,\ge\,
\left\lfloor\frac{1/2-P_1(n,D)}
{2\cdot P_2(n,D)}\right\rfloor\,-\,1,
\eqno(4.9)
$$
where}
$$
P_1(n,D)\eq\Pr\{S(\u,\tilde{\bar{\u}})\ge n-D\}\,=
\,q^{-n}\sum\limits_{t=0}^D\,
|\bar{\P}(n,n-t)|.\eqno(4.10)
$$
$$
P_2(n,D)\eq
\Pr\{S(\u,\v)\ge n-D\}\,=\,q^{-2n}\sum\limits_{t=0}^D\,
|\P(n,n-t)|,\eqno(4.11)
$$

\textbf{Proof of Lemma 4.1.}
Let
$\X=\{\x(1),\x(2),\dots,\x(2N)\}$ be an arbitrary DNA
 code of length $n$ and  size~$2N$.  Without
loss of generality, we put the codeword
$\x(N+k)\eq\tilde{\bar{\x}}(k)$ for any $k\in [N]$. In
virtue of this, code $\X$ satisfies the condition $(i)$ of
Definition~4.  Note that code $\X$ will satisfy the
condition $(ii)$ of Definition~4 if for an arbitrary pair
of codewords $(\x(k),\x(k'))$, $k\ne k'$, the
number
$$
S(\x(k),\x(k'))\le n-D-1.
$$
We will say that a  pair of codewords $(\x(k),\x(k+N))$,
$k=1,2,\dots,N$,
is an $D$-{\em bad pair in code} $\X$ if there exists a
codeword $\x(k')$ for which
$$
\mbox{either}\; S(\x(k),\x(k'))\ge n-D,\; k'\ne k,
\; \mbox{or}\; S(\x(k+N),\x(k'))\ge n-D,\;
k'\ne k+N.
$$
Otherwise, we will say that $(\x(k),\x(k+N))$,
$k=1,2,\dots,N$, is an $D$-{\em good pair in code}~$\X$.

Consider the {\em ensemble} of $q$-ary codes
$\X=\{\x(1),\x(2),\dots,\x(2N)\}$ of length $n$
and size $2N$, where codewords
$\x(1)$, $\x(2)$, $\dots$, $\x(N)$ are composed of $n\cdot N$
independent identically distributed letters having the
uniform distribution on~$\A$. One can easily understand that
for an arbitrary pair of random
codewords $(\x(k),\x(k'))$, $k\ne k'$, the distribution of
random variable $S(\x(k),\x(k'))$ has the form~(4.8).
Hence, using notations (4.10)-(4.11) and
the additive bound on the union probability, we have
$$
\Pr\{\mbox{pair}\;(\x(k),\x(k+N))\;
\mbox{is $D$-bad in code $\X$}\}\le
(2N-2)P_2(n,D)+P_1(n,D).
\eqno(4.12)
$$

Introduce the integer
$$
\tilde{N}\eq\left\lfloor\frac{1/2-P_1(n,D)}
{2\cdot P_2(n,D)}\right\rfloor\,+\,1.
$$
Inequality (4.12) means that for the ensemble of $q$-ary
codes $\X$ of length $n$ and size~$2\tilde{N}$,
$$
\Pr\{\mbox{pair}\;(\x(k),\x(k+\tilde{N}))\;
\mbox{is $D$-bad in code $\X$}\}\le\frac12,
\quad k=1,2,\dots,\tilde{N},
$$
i.e., for the given ensemble, the {\em average} number of
$D$-good pairs  $\ge \lfloor\tilde{N}/2\rfloor$.
Therefore, there exists an
$(n,D)$-code of
size~$\ge 2\lfloor\tilde{N}/2\rfloor\ge\tilde{N}-2$.
This yields~(4.10).

Lemma 4.1 is proved.
\medskip

For fixed parameter $u$, $0\le u\le 1$, define functions
$$
\p(u)\eq\varlimsup_{n\to\infty}\frac{\log_q|\P(n,\lceil (1-u)n\rceil)|}{n}
\quad\mbox{and}\quad
\bar{\p}(u)\eq\varlimsup_{n\to\infty}
\frac{\log_q|\bar{\P}(n,\lceil (1-u)n\rceil)|}{n}
$$
satisfying obvious inequalities $\;0\le\p(u)\le2\,$ and
$\,0\le\bar{\p}(u)\le1\,$.
One can easily understand that Lemma~4.1 yields
a {\em random coding bound} on the rate~(2.2) of $(n,\lfloor dn\rfloor)$-codes
which is given by
\medskip

\textbf{Lemma 4.2.}\quad
{\em Let $d$, $0<d<1$, be fixed. If}\quad
$\min\limits_{0\le u\le d}\,\{1-\bar{\p}(u)\}>0$,
\quad{\em then the rate}
$$
R_q(d)\,\ge\,\min\limits_{0\le u\le d}\,\{2-\p(u)\}.
$$
\medskip

If we  apply Lemmas~4.1 and~4.2 to a
similarity function $\S(\x,\y)$, then we need to investigate
the corresponding sets~(4.7). For instance, consider
the additive similarity $S^{\alpha}(\x,\y)$ which is defined as
the number of positions $i$, $i=1,2,\dots,n$,
where~$x_i=y_i$.  Let the corresponding sets~(4.7) be
$P^{\alpha}(n,s)$ and~$\bar{\P}^{\alpha}(n,s)$.  It is easy
to see that  the set $\bar{\P}^{\alpha}(n,s)$ is empty if
$s$ is odd. The sizes of sets $|P^{\alpha}(n,s)|$ and
$|\bar{\P}^{\alpha}(n,s)|$, $s=2,4,\dots$, are calculated
as follows:  $$ |\P^{\alpha}(n,s)|={n\choose
s}\,q^s\,q^{n-s}\,(q-1)^{n-s}=q^n\, {n\choose
s}\,(q-1)^{n-s}, $$ $$
|\bar{\P}^{\alpha}(n,s)|=|\P^{\alpha}(\lfloor
n/2\rfloor,s/2)|.
$$
Thus, for any $u$, $0<u<1$, the $\cap$-convex function
$$
\p^{\alpha}(u)\,\eq\,\varlimsup_{n\to\infty}
\frac{\log_q\left[q^n\,
{n\choose \lceil (1-u)n\rceil}\,(q-1)^{n-\lceil (1-u)n\rceil}\right]}{n}\,=
\,1+h_q(u)+u\log_q(q-1)
$$
and the $\cap$-convex function
$\;\bar{\p}^{\alpha}(u)=\p^{\alpha}(u)/2$. Obviously,
$$
\max\limits_{0\le u\le 1}\,\p^{\alpha}(u)\,=
\p^{\alpha}\left(\frac{q-1}{q}\right)\,=\,2.
$$
Therefore, if $0<d<\frac{q-1}{q}$, then
$$
\min\limits_{0\le u\le d}\,\{1-\bar{\p}^{\alpha}(u)\}=
\frac12\min\limits_{0\le u\le d}\,\{2-\p^{\alpha}(u)\}=
\frac12\left[1-h_q(d)-d\log_q(q-1)\right]>0.
$$
Hence, applying Lemma~4.2, we get the following
lower bound on  the rate $R^{\alpha}_q(d)$ of DNA codes based on the
additive similarity
$$
R^{\alpha}_q(d)\,\ge\, 1-h_q(d)-d\log_q(q-1),\quad
0<d<\frac{q-1}{q}.
$$
This bound  coincides  with the well-known Gilbert-Varshamov bound
on the rate of $q$-ary error-correcting codes  for the Hamming
metric~\cite{ms77}.

In Sect.~4.3 and~4.4, we will investigate the sizes of sets~(4.7) for
similarity functions  $\S^{\lambda}$ and~$\S^{\beta}$. Applying this
analysis, we will prove Theorems~4.1 and~4.2 with the help of Lemmas~4.1
and~4.2.


\subsection{Proof of Theorem 4.2}
\quad
Let  $s$, $\,1\le s\le n$, be an arbitrary integer and
$$
\P^{\beta}(n,s)\eq\{(\x,\y)\in\A^n\times\A^n\,:\,\S^{\beta}(\x,\y)=s\},
\quad
\bar{\P}^{\beta}(n,s)\eq\{\x\in\A^n\,:\,\S^{\beta}(\x,\tilde{\bar{\x}})=s\}
$$
denote  sets~(4.7) for similarity of blocks~$\S^{\beta}(\x,\y)$.

For a  fixed sequence $\z=(z_1,z_2,\dots,z_s)\in\A^s$, we
introduce the concept of its $j$-{\em block partition}
$$
\z=\{\b_1,\b_2,\dots,\b_{j-1},\b_j\},\qquad
j=1,2,\dots,\min\{s,\,n-s+1\},
\eqno(4.13)
$$
i.e., a  partition of $\z$ into $j$  nonempty blocks, where each
block contains  {\em consecutive} elements of~$\z$. Let
$\x=(x_1,x_2,\dots,x_n)\in \A^n$, be a fixed $q$-ary $n$-sequence.
Definition~2 means that a block partition $\z$ of the form (4.13) is
a block subsequence of $\x$ if $\z$ is a subsequence of $\x$,
i.e.,
$$
\z=\left(x_{i_1},x_{i_2},\dots,x_{i_{s-1}},x_{i_s}\right),\qquad
1\le i_1<i_2<\cdots<i_{s-1}<i_s\le n,
$$
and all blocks $\{\b_1,\b_2,\dots,\b_{j-1},\b_j\}$
consisting of  consecutive elements of the sequence $\x$
are  separated in~$\x$. In addition, if
a pair $\,(\x,\y)\in\P^{\beta}(n,s)\,$ (a sequence
$\x\in\bar{\P}^{\beta}(n,s))$, then there exists a block
partition $\z$  which is a  common block subsequence  between $\x$
and~$\y$ ($\x$ and~$\tilde{\bar{\x}}$), i.e., each of
sequences $\x$ and~$\y$ ($\x$ and~$\tilde{\bar{\x}}$)
contains  separated blocks $\{\b_1,\b_2,\dots,\b_{j-1},\b_j\}$
consisting of their consecutive elements.
\medskip

\textbf{Lemma 4.3.}
\quad {\em For any $s$, $1\le s\le n$, the size}
$$
|\P^{\beta}(n,s)|\le\,\,q^s\,\cdot
\sum\limits_{j=1}^{\min\{s,\,n-s+1\}}\,
{s-1\choose j-1}\cdot \left[q^{n-s}\cdot\,
{n-s+1\choose j}\right]^2.
\eqno(4.14)
$$
\medskip

\textbf{Proof of Lemma 4.3.} Let $M\ge1$ and $N\ge1$ be arbitrary integers.
For $M\ge N$, denote by $W_1(M\,;\,N)$ the number of all ways to distribute $M$
indistinguishable marbles in $N$ boxes provided that all $N$ boxes are
nonempty. Denote by $W_2(M\,;\,N)$ the number of all ways to distribute $M$
indistinguishable marbles in $N$ boxes if empty boxes are accepted.
It is well-known that
$$
\quad W_1\left(M\,;\,N\right)={M-1\choose N-1},\;\; M\ge N, \quad\mbox{and}\quad
\quad W_2(M\,;\,N)={M+N-1\choose N-1}.
$$
Obviously, for any  $\z\in\A^s$, the number of all its
$j$-block partitions of the form (4.13) is
$$
 W_1(s\,;\,j)={s-1\choose j-1},\qquad
j=1,2,\dots,\min\{s,\,n-s+1\}.
\eqno(4.15)
$$
If $M=(n-s)-(j-1)$ and $N=j+1$, then we have $N+M-1=n-s+1$, $N-1=j$
and
$$
W_2\left((n-s)-(j-1)\,;\,j+1\right)\,=\,{n-s+1\choose j}
\eqno(4.16)
$$
is an upper bound on the cardinality  of the
following set of $q$-ary $n$-sequences.
These $n$-sequences are obtained by $M=(n-s)-(j-1)$ insertions
of a fixed $M$-collection of $q$-ary letters (marbles) into $N=j+1$
"spaces"  generated by a fixed  $q$-ary $s$-sequence $\z$ having a
fixed $j$-block partition (4.13), namely: the space before $\b_1\,$, the space after
$\b_j\,$ and $j-1$ inter-block  spaces of (4.13) which are marked by a fixed
$(j-1)$-collection of separating $q$-ary letters (marbles).
The given interpretation of formulas (4.15)-(4.16) leads to~(4.14).

Lemma 4.3 is proved.

For any fixed sequence $\z\in\A^s$ and
its $j$-block partition (4.13), we introduce a
{\em reverse complement} $j$-block partition
$$
\tilde{\bar{\z}}\eq
\{\tilde{\bar{\b}}_j,\tilde{\bar{\b}}_{j-1},\dots,\tilde{\bar{\b}}_2,
\tilde{\bar{\b}}_1\},\qquad
j=1,2,\dots,\min\{s,\,n-s+1\}.
$$

\textbf{Lemma 4.4.}\quad
{\em The set $\bar{\P}^{\beta}(n,s)$ is empty if $s\ge1$ is odd.
If $s\ge2$ is even and an $n$-sequence $\x\in\bar{\P}^{\beta}(n,s)$,
then there
exist an integer $j$, $j=1,2,\dots,\min\{s,\,n-s+1\}$ and
a self-reverse complementary $s$-sequence
$\z=\tilde{\bar{\z}}$, $|\z|=s$,
of the form $(4.13)$ which is a common block
subsequence between $\x$ and~$\tilde{\bar{\x}}$ and $\z$ has
a self reverse complementary block partition
$$
\z=\{\b_1,\b_2,\dots,\b_{j-1},\b_j\}=
\{\tilde{\bar{\b}}_j,\tilde{\bar{\b}}_{j-1},\dots,\tilde{\bar{\b}}_2,
\tilde{\bar{\b}}_1\}=\tilde{\bar{\z}},
$$
i.e., block $\b_1=\tilde{\bar{\b}}_j$, block $\b_2=\tilde{\bar{\b}}_{j-1}$,
$\dots$,
block $\b_{j-1}=\tilde{\bar{\b}}_2$, and block $\b_{j}=\tilde{\bar{\b}}_1$}.

\textbf{Proof of Lemma 4.4.} Consider an arbitrary
$\x\in\bar{\P}^{\beta}(n,s)$ and its reverse complement
$\tilde{\bar{\x}}$. Let a sequence $\z\in\A^m$, $m\in [s]$,
be a block subsequence (BSS) of $\x$. Then one can easily
see that $\z$ is a BSS of $\tilde{\bar{\x}}$ if and only
if its reverse complement $\tilde{\bar{\z}}$ is a BSS of~$\x$.
This means that the following two statements are
equivalent.
\begin{enumerate}
\item
  The set $\bar{\P}^{\beta}(n,s)$ is empty if $s$ is odd.
If $s$ is even and a block partition $\z$, $|\z|=s$, is a common BSS
between $\x$ and~$\tilde{\bar{\x}}$, then there exists
a sequence $\z'=\tilde{\bar{\z'}}$ of length $|\z'|=|\z|=s$ having a
self-reverse complementary block partition $\z'$ which is a
common BSS between $\x$ and~$\tilde{\bar{\x}}$.
\item
 The set $\bar{\P}^{\beta}(n,s)$ is empty if $s$ is odd.
If $s$ is even and  block partitions $\z$, $\tilde{\bar{\z}}$
of length $|\z|=|\tilde{\bar{\z}}|=s$ are BSS of
$\x$, then there exists
a sequence $\z'=\tilde{\bar{\z'}}$ of length $|\z'|=|\z|=s$ having a
self-reverse complementary block partition $\z'$ which is a
BSS of~$\x$.
\end{enumerate}

Obviously, statement~1 is equivalent to the statement of Lemma~4.4.
Hence, to complete the proof of Lemma~4.4, we need to check statement~2.
For any $s\in[n]$, one can write
$$
\z=\left(x_{i_1},x_{i_2},\dots,x_{i_{s-1}},x_{i_s}\right),\qquad
1\le i_1<i_2<\cdots<i_{s-1}<i_s\le n,
$$
and
$$
\tilde{\bar{\z}}=\left(x_{k_1},x_{k_2},\dots,x_{k_{s-1}},x_{k_s}\right),\qquad
1\le k_1<k_2<\cdots<k_{s-1}<k_s\le n,
$$
where
$$
x_{k_1}=\bar{x}_{i_s},\quad x_{k_2}=\bar{x}_{i_{s-1}},\quad
\dots,\quad x_{k_{s-1}}=\bar{x}_{i_2},\quad x_{k_s}=\bar{x}_{i_1}.
\eqno(4.17)
$$

Let $s\ge1$ be an odd integer. From $(4.17)$ it follows
 $x_{k_{\lceil s/2\rceil}}=\bar{x}_{i_{\lceil s/2\rceil}}$.
Hence, $i_{\lceil s/2\rceil}\ne k_{\lceil s/2\rceil}$
because for any element $x\in\A=\{0,1,\dots,q-1\}$,
$q=2,4,\dots$, its complement $\bar{x}\eq(q-1)-x\ne x$.
Without loss of generality, we say
$i_{\lceil s/2\rceil}< k_{\lceil s/2\rceil}$. Then, in
virtue of $(4.17)$, the $q$-ary sequence
$$
\z'\eq\left(x_{i_1},x_{i_2},\dots,x_{i_{\lceil s/2\rceil}},
x_{k_{\lceil s/2\rceil}},\dots,x_{k_{s-1}},x_{k_s}\right)
$$
of length $\lceil s/2\rceil+\lceil s/2\rceil=s+1$ is a
self-reverse complementary common BSS between~$\x$ and~$\tilde{\bar{\x}}$.
This contradicts to the condition $\x\in\bar{\P}^{\beta}(n,s)$,
i.e., the set $\bar{\P}^{\beta}(n,s)$ is empty if $s$ is odd.

Let $s=2t$, $t=1,2,\dots$ be an even integer.
Without loss of generality, we say
$$
\frac{i_t+i_{t+1}}{2}\le\frac{k_t+k_{t+1}}{2},
\quad\mbox{i.e.,}\quad
i_t<k_{t+1} \quad\mbox{because}\quad i_t<i_{t+1}
\quad\mbox{and}\quad k_t<k_{t+1}.
$$
Then, in virtue of $(4.17)$, the
$q$-ary sequence
$
\z'\eq\left(x_{i_1},x_{i_2},\dots,x_{i_t},
x_{k_{t+1}},\dots,x_{k_{s-1}},x_{k_{s}}\right)
$
of length $s=2t$ is a self-reverse complementary  BSS of~$\x$.

Statement~2 and Lemma 4.4 are proved.

Lemma 4.4 and the arguments used for Lemma~4.3 lead to

\textbf{Lemma 4.5.}\quad {\em For any even $s$, $s\in [n]$, the
size}
$$
|\bar{\P}^{\beta}(n,s)|\le\,\,q^{s/2}\,\cdot
\sum\limits_{j=1}^{\min\{s,\,n-s+1\}}\,
{s/2-1\choose \lceil j/2\rceil-1}\,\cdot\,\left[q^{n-s}
{n-s+1\choose j}\right].
$$

For $s\in [n]$, consider numbers
$$
B(n,s)\,\eq\,\max\limits_{1\le j\le\min\{s,\,n-s+1\}}\;
\left\{{s-1\choose j-1}\,\cdot
\,{n-s+1\choose j}^2\right\}.\eqno(4.18)
$$
\medskip

\textbf{Proof of Statement $(i)$ of Theorem 4.2.}\quad
If $n\to\infty$, $k=0,1,2,\dots$
is fixed and $s=n-k$, then the maximum in (4.18) is asymptotically
achieved at $j=n-s+1=k+1$ and the maximal value
$$
B(n,n-k)=\frac{n^k}{k!}\cdot(1+o(1)).
$$
Hence,  Lemma~4.3 yields the asymptotic
inequality
$$
|\P^{\beta}(n,n-k)|\le
q^{2n}\cdot\frac{n^k\,q^k}{k!\,q^{n}}\cdot(1+o(1)),\quad
n\to\infty,\quad k=0,1,2,\dots.
$$
If $n\to\infty$ and $D=1,2,\dots$ is fixed, then
definition~(4.11)  means that
$$
P_2(n,D)
\,\eq\,q^{-2n}\sum\limits_{k=0}^D\,
|\P^{\beta}(n,n-k)|\,\le\,\frac{n^D\,q^D}{D!\,q^{n}}\cdot(1+o(1)).
$$
The similar arguments using  Lemma~4.5 and  definition~(4.10)
 show that
$$
P_1(n,D)
\,\eq\,q^{-n}\sum\limits_{k=0}^D\,
|\bar{\P}^{\beta}(n,n-k)|\,=\,o(1).
$$
Thus, Lemma~4.1 yields (4.5).

Statement $(i)$ of Theorem~4.2 is proved.
\medskip

\textbf{Proof of Statement $(ii)$ of Theorem 4.2.}\quad
Let $u$, $0<u<1$, be fixed parameter. Define the function
$$
E_q(u)\eq\lim\limits_{n\to\infty}
\frac{\log_q\, B\left(\,n,\,\lceil(1-u)n\rceil\right)}{n},\qquad 0<u<1.
$$
Lemmas~4.3 and~4.5 yield  upper bounds on
functions $\p^{\beta}(u)$ and $\bar{\p}^{\beta}(u)$ used in Lemma~4.2:
 $$
 \p^{\beta}(u)\eq\varlimsup\limits_{n\to\infty}\,
 \frac{\log_q |\P^{\beta}(n,\lceil (1-u)n\rceil)|}{n}\,\le\,
 (1+u)+E_q(u),
 $$
$$
 \bar{\p}^{\beta}(u)\eq\varlimsup\limits_{n\to\infty}\,
 \frac{\log_q |\bar{\P}^{\beta}(n,\lceil (1-u)n\rceil)|}{n}\,\le\,
\frac12\left[(1+u)+E_q(u)\right].
 $$
Therefore, Lemma~4.2  gives a random coding bound on the rate $R_q^{\beta}(d)$ of
$q$-ary  DNA $(n,\lfloor dn\rfloor)$-codes based on the block
similarity.
One can easily check that the given lower bound  $\underline{R}_q^{\beta}(d)$
can be written in the form
$$
R_q^{\beta}(d)\,\ge\,\underline{R}_q^{\beta}(d)\,=\, (1-d)-E_q(d),\qquad
E_q(d)\,=\,\max\limits_{0\le v\le d}\;F_q(v,d),\eqno(4.19)
$$
where
$$
F_q(v,d)\,\eq\,(1-d)\,h_q\left(\frac{v}{1-d}\right)\,+
\,2d\,h_q\left(\frac{v}{d}\right).
$$
The derivative of the binary entropy function $h_q(v)$ is
$$
h'_q(v)=\log_q\frac{1-v}{v},\qquad  0<v<1.
$$
Thus, the partial derivative of the function  $F_q(v,d)$ is
\smallskip
$$
\frac{\partial F_q(v,d)}{\partial v}\,=\,
\log_q\frac{(1-d)-v}{v}\,+\,2\log_q\frac{d-v}{v}\,=\,
\log_q\frac{[(1-d)-v](d-v)^2}{v^3}
$$
and for a fixed $d$, $0<d<1/2$,  equation $\partial F_q(v,d)/\partial v\,=\,0$ is
equivalent to equation~(4.3).
The binary entropy function $h_q(v)$ is $\cap$-convex function
of parameter $v$, $0<v<1$.
Hence, formulas (4.3)-(4.4) give the solution of the maximization
problem (4.19) for $\cap$-convex function $F_q(v,d)$ of parameter
$v$, $0\le v\le d$. This yields~(4.6).

Theorem 4.2 is proved.

\subsection{Proof of Theorem 4.1}
\quad
Let  $s$,\quad $0\le s\le n$, be an arbitrary integer and
$$
\P^{\lambda}(n,s)\eq\{(\x,\y)\,:\,S^{\lambda}(\x,\y)=s\}, \quad
\bar{\P}^{\lambda}(n,s)\eq\{\x\,:\,S^{\lambda}(\x,\tilde{\bar{\x}})=s\}
$$
denote the sets from Lemma~4.1 for the deletion similarity. An
upper bound on the size $|\P^{\lambda}(n,s)|$ is based on the
following well-known~\cite{lev74,lev01} result.
\medskip

\textbf{Lemma 4.6.}~\cite{lev74,lev01}. {\em Let $n$ and $s$ be
integers, $0\le s\le n$. For an arbitrary sequence
$\y\in\A^s$ denote by $\B_q(\y,n)$ the set of all sequences
$\x\in\A^n$ that include $\y$ as a subsequence, i.e., that can be
obtained from $\y$ by $n-s$ insertions. Then for the fixed $n$ and
$s$, the size of $\B_q(\y,n)$ does not depend on $\y$ and has the
form}
$$
|\B_q(\y,n)|=\sum_{k=0}^{n-s}{n\choose k}(q-1)^k\eq B_q(n,s).
\eqno(4.20)
$$

\textbf{Proof of Lemma~4.6}.
We will use
the induction over~$s$. For $s=0$ and
$s=1$, Lemma~4.4  is trivial. Assume that Lemma~4.4 is proved
for all integers less than~$s\ge2$.
Consider  an arbitrary $s$-sequence~$\y=(y_1,y_2,\ldots,y_s)$ and
its  $(s-1)$-subsequence $\y'\eq(y_2,y_3,\ldots,y_s)$. Divide
the set $\B_q(\y,n)$ into the sum of mutually disjoint sets
$\B_q^k(\y,n)$, $k=1,2,\ldots,n-s+1$, where the set
$\B_q^k(\y,n)$ is composed of $n$-sequences $\x=(x_1,x_2,\dots,x_n)\in\B_q(\y,n)$
such that $x_i\ne y_1$ for $i=1,2,\ldots,k-1$ and $x_k=y_1$.
Obviously, any such sequence $\x$ belongs to the
set $\B_q(\y,n)$ if and only if the $(n-k)$-sequence
$(x_{k+1},x_{k+2},\ldots,x_n)$ contains~$\y'$. In virtue of
the induction hypothesis, the size
$$
|\B_q^k(\y,n)|=(q-1)^{k-1}|\B_q(\y',n-k)|=
(q-1)^{k-1}B_q(n-k,s-1),
$$
i.e., for any $k=1,2,\ldots,n-s+1$, the size
$|\B_q^k(\y,n)|$ is the same for all $s$-sequences~$\y$.
This means that the size $|\B_q(\y,n)|$  does not depend
on~$\y$ as well. To complete the proof, we consider the $s$-sequence
$\y=(0,0,\ldots,0)$ for which the equality of Lemma~4.4 is trivial.

Lemma~4.6 is proved.


\textbf{Lemma 4.7.}
{\em The set $\bar{\P}^{\lambda}(n,s)$ is empty if $s$ is odd.
If $s$ is an even  number and a sequence $\x\in\bar{\P}^{\lambda}(n,s)$,
then there exists a self reverse complementary sequence
$\z=\tilde{\bar{\z}}$, $|\z|=s$, which is a common
subsequence between $\x$ and $\tilde{\bar{\x}}$}.

The proof of Lemma~4.7 is omitted here because it can be easily
obtained by an evident modification of our arguments used
for Lemma~4.4.

Lemmas~4.6 and~4.7  yield
$$
|\P^{\lambda}(n,s)|\le\,\,q^s\,\cdot\,\left[B_q(n,s)\right]^2,
\quad
|\bar{\P}^{\lambda}(n,s)|\le\,\,q^{s/2}\,\cdot\,B_q(n,s),
\quad 0\le s\le n.
\eqno(4.21)
$$
Applying (4.20)-(4.21), Lemma~4.1 and arguments for
Statement~$(i)$ of Theorem~4.2, one can easily prove (4.1), i.e.,
 Statement~$(i)$ of Theorem~4.1.

If $u$, $\;0\le u\le (q-1)/q$, is  fixed, then
from definition (4.20) it follows
$$
\lim\limits_{n\to\infty}\,\frac{\log_q B_q(n,\lceil
(1-u)n\rceil)}{n}\,=\,
u\log_q(q-1)+h_q(u).
$$
Therefore, applying (4.21), we have
$$
\p^{\lambda}(u)\eq\varlimsup_{n\to\infty}
\frac{\log_q|\P^{\lambda}(n,\lceil (1-u)n\rceil)|}{n}\le
1-u+2u\log_q(q-1)+2h_q(u)
\eqno(4.22)
$$
and
$$
\bar{\p}^{\lambda}(u)\eq\varlimsup_{n\to\infty}
\frac{\log_q|\bar{\P}^{\lambda}(n,\lceil (1-u)n\rceil)|}{n}\,
\le\,\frac12\cdot[1-u+2u\log_q(q-1)+2h_q(u)],
\eqno(4.23)
$$
provided that $0<u\le(q-1)/q$.  Hence, if $0<d<(q-1)/q$,
then from (4.22)-(4.23) it follows
$$
\min\limits_{0\le u\le d}\,\{1-\bar{\p}^{\lambda}(u)\}\ge
\frac12\cdot[1+d-2d\log_q(q-1)-2h_q(d)],\eqno(4.24)
$$
$$
\min\limits_{0\le u\le d}\,\{2-\p^{\lambda}(u)\}\ge
1+d-2d\log_q(q-1)-2h_q(d).\eqno(4.25)
$$
Inequalities (4.24)-(4.25) and Lemma~4.2 yield~(4.2), i.e.,
 Statement~$(ii)$ of Theorem~4.1.

Theorem~4.1 is proved.

\end{document}